\documentclass[10pt,aps,pre,onecolumn,superscriptaddress,nofootinbib]{revtex4-2}

\usepackage{amsmath,amssymb,graphicx,bm}
\usepackage{hyperref}
\usepackage{xcolor}%
\usepackage[percent]{overpic}
\usepackage{epigraph}
\usepackage{verbatim}

\begin{document}

\title{To clean or not to clean: The free‑rider problem in sequentially shared resources}

\author{Alexander Feigel}
\email{alexander.feigel@mail.huji.ac.il}
\affiliation{The Racah Institute of Physics, The Hebrew University of Jerusalem, Edmond J. Safra Campus, Jerusalem 9190401, Israel}

\author{Alexandre V. Morozov}
\email{morozov@physics.rutgers.edu}
\affiliation{Department of Physics and Astronomy, Rutgers, The State University of New Jersey, 136 Frelinghuysen Road, Piscataway, New Jersey 08854, USA}

\date{\today}

\begin{abstract}
Shared resources enhance productivity yet at the same time provide channels for biological and digital contamination, turning physical or digital hygiene into a cooperation dilemma prone to free-riding. Here we introduce a game of sequential sharing of common resources, an empirically parameterized evolutionary model of population dynamics in
sequential-use settings such as gyms and shared
workspaces. The success of the strategies implemented in the model,
such as cleaning equipment before or after use, are based on the
trade-offs between cleaning costs, contamination risk, and social
incentives to mitigate disease transmission. We find that cooperative
hygiene can be achieved by lowering the effective costs of
cleaning, strengthening pro-social incentives, and monitoring
population-level noncompliance.
Remarkably, stability of fully altruistic populations is primarily affected by the cleaning costs.
In contrast, increasing effective infection costs, for example through punishment, appears less important in this case.
The model's evolutionary dynamics exhibit multi-stability,
hysteresis, and abrupt shifts in strategy composition, broadly
consistent with empirical observations from shared-use facilities. Our
framework offers testable predictions and is amenable to quantitative
calibration with behavioral and environmental data. Our predictions
can be used to inform the design of cost-effective public health and digital security policies.
\end{abstract}

\maketitle


\section{Introduction}

Physics is increasingly applied to studies of social behavior. Methods from non-linear dynamical systems and network science are frequently employed to understand, predict, and guide interventions in animal and human social dynamics~\cite{Szabo2007,Castellano2009,Nguyen2025}. There are major
multidisciplinary efforts aimed at advancing theory and modeling of
social behaviors~\cite{Baer1968,Sandholm2010a,Hofbauer1998a} such as cooperation~\cite{Axelrod1981,Nowak2006a}.
In particular, shared use of equipment poses a game-theoretic dilemma: individuals are tempted to reduce personal contamination risk by cleaning equipment before use and avoiding the cost of cleaning after use, with some people avoiding cleaning altogether, while the population as a whole has a collective interest in maintaining safe, disease-free environment. Shared equipment poses a considerable risk of infection~\cite{Markley2012,Kiborus2025} and its associated costs, such as missed work and medical expenses. 

What are the factors that promote and sustain high population-level rates of cleaning, particularly altruistic post-use cleaning? An experimental study found that combining social pressure (signage) with easy access to cleaning supplies significantly increased cleaning of equipment after use in a university gym, although some noncompliance persisted~\cite{Elba2018}. Similarly, informative signage promotes recycling~\cite{Austin1993} and convenient access to hygiene resources increases hand-washing~\cite{Fournier2012}. Experiences during the COVID-19 epidemic have shown that epidemiological costs strongly shape personal protective behaviors across populations~\cite{Shawler2021}. Taken together, these observations indicate that cleaning practices can be modeled quantitatively as functions of social pressure driven by population-level interest in reducing both overall contamination levels and infection costs.

Cooperative cleanliness has deep historical roots, from the earliest communal lodgings to contemporary shared environments such as co-working spaces, public transportation, and exercise facilities. It now extends to the digital realm, including shared digital storage and repositories used to collaborate on software projects and presentations. Sharing tools or digital resources poses risks of contamination, including germs, viruses, and malware. Given its importance across physical and digital spaces, cooperative cleanliness is sustained by social norms and regulatory oversight aimed at shaping individual behavior.

Modeling human behavior is challenging because it involves complex phenomena such as learning, imitation, and responses to environmental and social pressures. Models of social behavior often rely on the replicator equation framework that adapts Darwinian selection to game-theoretical settings~\cite{Smith1973,Hofbauer1998a,Nowak2006book}, with imitation based on beliefs and observations playing the role of genetic inheritance.
Replicator dynamics captures individual, payoff-driven incentives and provides a standard framework for modeling behavioral adjustment through imitation, social learning, and reinforcement.

Here we develop a model of interactions with shared equipment
used sequentially by multiple users, for example in gyms or co-working spaces, in the presence of infection in the population as a whole. Infected users contaminate equipment, and uninfected users can become infected by
using it. Individual strategies are shaped by personal incentives to
reduce infection risks and by the social pressure to lower
overall contamination levels.
We model imitation of successful strategies using replicator equation dynamics, which promotes strategies with above-average payoffs. We assume no mutations or exogenous noise, so that new strategies do not appear spontaneously. We also assume that adjustments in population frequencies are sufficiently slow to justify a continuous-time approximation. The imitation process is governed by two primary costs: the cost of cleaning and the cost of infection.

To model social pressure, we adopt a gradient approach in which the driving force is analogous to a physical force in a potential field~\cite{Friedman2010}, with the potential equal to the total level of contamination. In this approach, individuals perceive contamination levels through signage, news, and other forms of communication, and subsequently adjust their behavior in order to reduce contamination risks.
Potential games provide an established framework for modeling social pressure~\cite{Rosenthal1973,Monderer1996,Sandholm2010a}.

We find that if individuals imitate one another and respond to
social pressure to reduce overall contamination,
altruistic post-usage cleaning is sustained by social pressure
and by lowering cleaning costs. 
As the perceived cost of infection increases, the
collective response exhibits an abrupt transition to near-total
protection. A similarly abrupt transition is predicted when access to cleaning resources improves and cleaning costs decline. These results are broadly consistent with existing empirical findings and form a basis for further experimental and observational studies of human societies.

\section{Model}

We consider a setting where individuals use shared equipment
sequentially, as in a gym. Infected users contaminate any equipment
they touch; susceptible users become infected if they subsequently
use contaminated equipment. We assume that pre-use cleaning removes all existing contamination; post-use cleaning completely removes contamination deposited by the current and previous users.

Let $\rho_A, \rho_B, \rho_C, \rho_D$ denote the population frequencies of strategies A, B, C, and D, respectively:
\begin{itemize}
\item \textbf{A:} cleans After use (purely altruistic strategy)
\item \textbf{B:} cleans Before use (purely selfish strategy)
\item \textbf{C:} never cleans (Cheater; a strategy that disregards the cleaning norms)
\item \textbf{D:} cleans both before and after use (a Dutiful citizen who follows the cleaning norms)
\end{itemize}

Uppercase letters denote strategy names. The frequencies are normalized:
\begin{equation} \label{eq:rho_sum}
    \sum_{i \in \{A,B,C,D\}} \rho_i = 1.
\end{equation}

The fitness of each strategy is
\begin{eqnarray} \label{eq:fD_unique1}
f_A &=& W_{\text{cln}} + W_{\text{inf}} \, I_A, \\
f_B &=& W_{\text{cln}}, \nonumber \\
f_C &=& W_{\text{inf}} \, I_C, \nonumber \\
f_D &=& 2 W_{\text{cln}}, \nonumber
\end{eqnarray}
where \(W_{\text{cln}}<0\) is the cost of protection and \(W_{\text{inf}}<0\) is the cost of infection. Strategies A and C face equipment-mediated infection with probabilities \(I_A\) and \(I_C\), respectively; B and D do not since they clean the equipment before using it,
with strategy D paying twice the protection cost.
Users following $A$ or $C$ strategies may
encounter contamination left by infected $B$ or $C$
users. Because they share this exposure pathway, their
infection risks per single instance of using the equipment are identical (see Appendix~\ref{sec:deriv_indiv_prob} for details):
\begin{equation} \label{eq:currents}
I_A = I_C = I(\rho_B, \rho_C; x),
\end{equation}
where
\begin{equation} \label{eq:I_rhoB_rhoC_x}
I(\rho_B, \rho_C; x) = \frac{x(1 - x)(\rho_B + \rho_C)}{1 - \rho_C(1 - x)}.
\end{equation}
Here, $I$ is the per-use infection probability (i.e., the probability
of becoming infected after a single instance of equipment usage);
$x \in [0,1]$ denotes infection prevalence in the general population (probability that an arriving user is already infected). The parameter $x$ is common to all the strategies under the assumption that most transmission events occur outside the gym.

Equation~\eqref{eq:I_rhoB_rhoC_x} depends differently on $\rho_{B}$ and $\rho_{C}$ because type $C$ can form chains of contamination transmission. 
In a population with a small amount of $C$, $B$ is the main driver of contamination spread:
\begin{equation} \label{eq:Iapprox:1}
I \approx x(1 - x)\rho_B
\end{equation}
which follows from Eq.~\eqref{eq:I_rhoB_rhoC_x} when $\rho_{C} \ll 1$. In
the $\rho_{B} \ll 1,~\rho_{C} \approx 1$ limit, Eq.~\eqref{eq:I_rhoB_rhoC_x} yields:
\begin{equation} \label{eq:Iapprox:2}
I \approx (1 - x)\rho_C.
\end{equation}
These contrasting limits show that when $B$ dominates,
transmission requires encounters between contaminated and
uncontaminated individuals (hence the factor $x(1-x)$), whereas when
$C$ dominates, contamination approaches the probability that a
non-contaminated individual uses the equipment (proportional to
$1-x$).

Social pressure seeks to minimize the total number of
infections~\cite{Bauch2003}. Accordingly, we introduce the social pressure potential as
\begin{equation} \label{eq:phi_inf}
\sigma \Phi_{\text{soc}} = I(\rho_{B},\rho_{C},x)\,(\rho_A + \rho_C),
\end{equation}
where $\Phi_{\text{soc}}$ is proportional to the total number of infections and the constant $\sigma \ge 0$ quantifies the strength of social pressure (see Appendix~\ref{sec:deriv_indiv_prob} for details). 
Here, $I(\rho_{B},\rho_{C},x)$ (Eq.~\eqref{eq:I_rhoB_rhoC_x}) denotes an individual's probability of infection from using the equipment, equally applicable to strategies $A$ and $C$. Consequently, $\Phi_{\text{soc}}$ depends explicitly on $\rho_A$, $\rho_B$ and $\rho_C$, and implicitly on $\rho_D$ via the constraint in Eq.~\eqref{eq:rho_sum}. For example, increasing $\rho_D$ reduces the shares of the other strategies and thus changes Eq.~\eqref{eq:phi_inf}.

Each strategy evolves in time under two forces: replicator dynamics that takes relative fitness of each strategy into account and social pressure given by the gradient of the social pressure potential $\Phi_{\text{soc}}$:
\begin{eqnarray} \label{eq:drhoD_uniqueu}
\dot{\rho}_A &=& \rho_A (f_A -
                 \bar{f}) - \sigma \frac{\partial \Phi_{\text{soc}}}{\partial \rho_{A}},
                   \\
\dot{\rho}_B &=& \rho_B (f_B -
                 \bar{f}) - \sigma \frac{\partial \Phi_{\text{soc}}}{\partial \rho_{B}}, \nonumber \\
\dot{\rho}_C &=& \rho_C (f_C -
                 \bar{f}) - \sigma \frac{\partial \Phi_{\text{soc}}}{\partial \rho_{C}}, \nonumber \\
\dot{\rho}_D &=& \rho_D (f_D -
\bar{f}) - \sigma \frac{\partial \Phi_{\text{soc}}}{\partial \rho_{D}}. \nonumber  
\end{eqnarray}
Here, \(\bar{f}\) denotes the average fitness:
\begin{equation} \label{eq:avgf_unique1}
\bar{f} = \sum_{i \in \{A,B,C,D\}} \rho_i f_i.
\end{equation}
The dot indicates a derivative with respect to time: $\dot{f}(t) = df(t)/dt$.

The above system of nonlinear equations governs the dynamics in the three-dimensional state space \((\rho_A, \rho_B, \rho_D)\), with \(\rho_C = 1 - \rho_A - \rho_B - \rho_D\). The parameters in Eq.~\eqref{eq:drhoD_uniqueu} can be reduced by rescaling time ($t \to \sigma t$): only the ratios \(W_{\text{cln}}/\sigma\) and \(W_{\text{inf}}/\sigma\) appear in the rescaled equations, allowing us to set \(\sigma = 1\) without loss of generality. The dynamics then depend on two dimensionless parameters, \(W_{\text{cln}}\) and \(W_{\text{inf}}\), which measure the costs of protection and infection relative to the strength of social pressure.
The model in Eq.~\eqref{eq:drhoD_uniqueu} is well suited to numerical analysis and admits approximate analytical treatment in some cases. Note that, unlike the replicator terms which are automatically zero when any $\rho_i$ reaches $0$ or $1$, the social pressure terms are generally non-zero at the limits of allowed population frequencies and therefore need to be regularized explicitly: accordingly, we set ${\partial \Phi_{\text{soc}}} / {\partial \rho_{i}} \big|_{\rho_i = \{0,1\}} = 0$ to prevent population frequencies from leaving the physically admissible $[0,1]$ range.


In the remainder of this work, we often set $x =0.5$, which corresponds to the maximum infection probability in Eq.~\eqref{eq:I_rhoB_rhoC_x}.
As discussed in Appendix~\ref{sec:deriv_indiv_prob}, the effect of $x$ on the evolutionary dynamics and the structure of fixed points is expected to be minimal for most parameter combinations.
Furthermore, we often reduce the dimension of the model by excluding
`dutiful citizens' (i.e., by setting $\rho_{D}=0$).
Strategy $D$ is strictly dominated by $B$
for any $W_{\text{cln}}>0$ in replicator dynamics because $f_D < f_B$ in Eq.~\eqref{eq:fD_unique1}; thus, we generally expect $\rho_D$ to be small.

\section{Results}

\begin{figure}[t]
  \centering
  \begin{minipage}[b]{0.48\textwidth}
    \begin{overpic}[width=\textwidth]{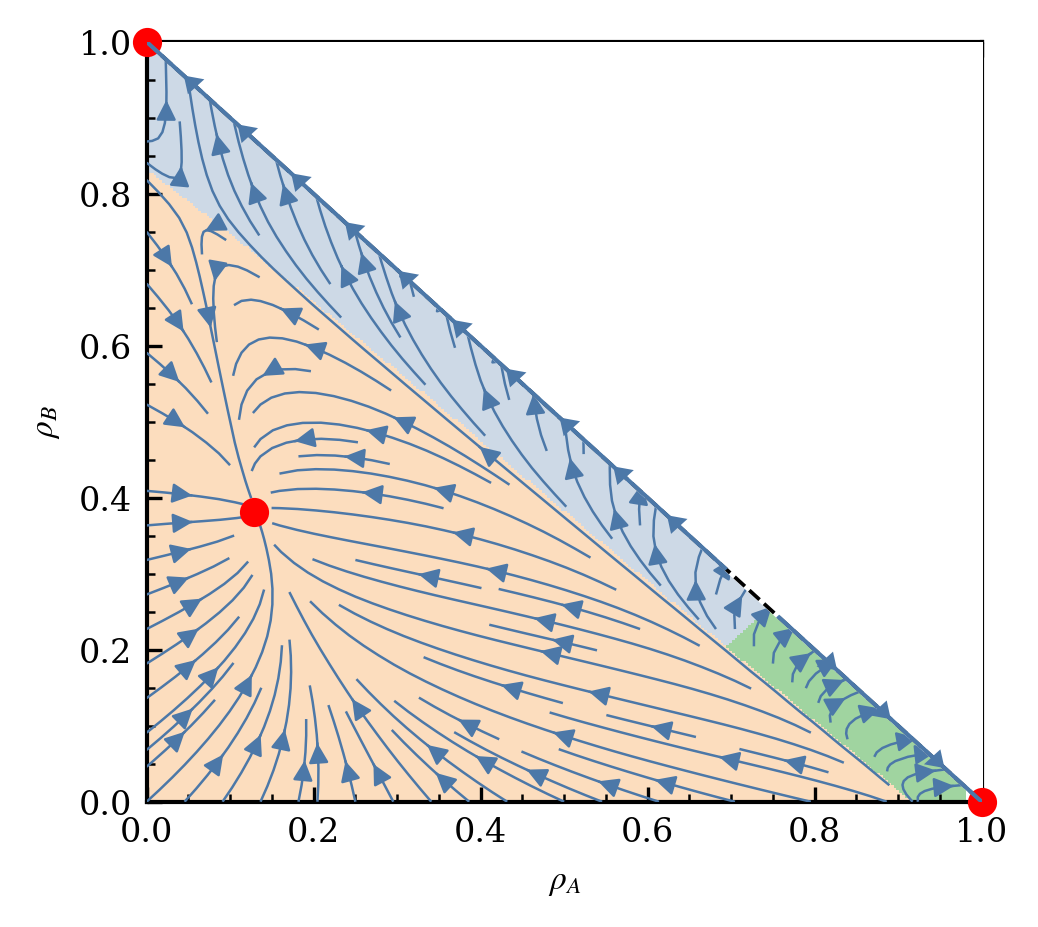}
            \put(4,94){\makebox(0,0)[l]{\footnotesize (a)}}   
    \end{overpic}
  \end{minipage}\hfill
  \begin{minipage}[b]{0.48\textwidth}
    \begin{overpic}[width=\textwidth]{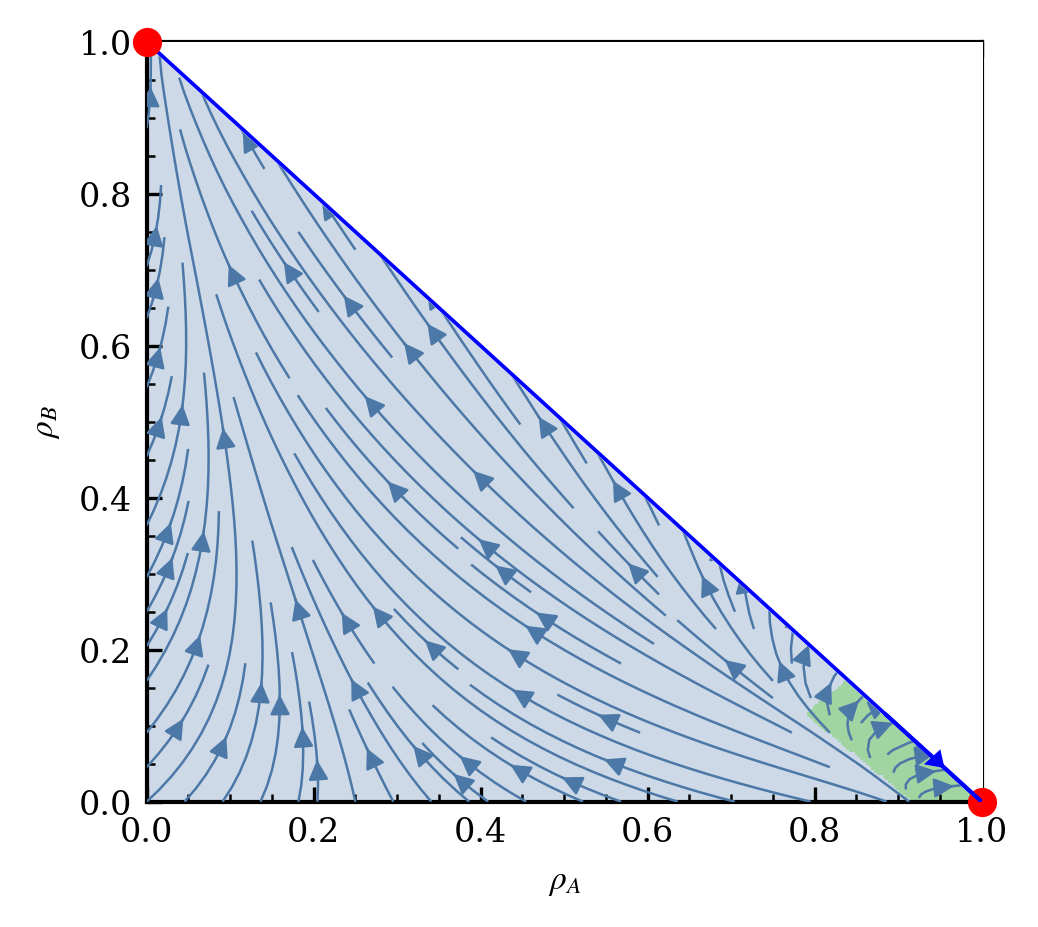}
            \put(4,94){\makebox(0,0)[l]{\footnotesize (b)}}   
    \end{overpic}
  \end{minipage}
  \caption{
    \textbf{Numerical analysis of stable states and their basins of attraction.} Shown are vector fields over the strategy frequency space
    $(\rho_A,\rho_B)$, assuming that $\rho_{D}=0$ and hence $\rho_{C}=1-\rho_{A}-\rho_{B}$. Blue arrows depict the flows driven by selective forces and social pressure (Eq.~\eqref{eq:drhoD_uniqueu}). Red dots mark stable equilibria.
    (a) Low infection cost: $W_{\text{cln}}=-4$, $W_{\text{inf}}=-5$.
    There are three stable equilibria: an altruistic monomorphic $A$ state ($\rho_{A}^*=1$), a selfish monomorphic
    $B$ state ($\rho_{B}^*=1$), and a mixed state comprising strategies
    $A$, $B$, and $C$. Background colors (green, blue, coral) denote numerically mapped regions of attraction of the altruistic, selfish,
    and mixed states, respectively. Most initial conditions converge to the mixed state.
    (b) High infection cost: $W_{\text{cln}}=-4$, $W_{\text{inf}}=-15$.
    There are two stable equilibria: an altruistic monomorphic $A$ state ($\rho_{A}^*=1$) and a selfish monomorphic $B$ state ($\rho_{B}^*=1$).
    The mixed state becomes unstable in this regime and cheaters $C$ go extinct.
    Background colors (green, blue) denote numerically mapped regions of attraction of the altruistic and selfish states, respectively. Note that most initial conditions converge to the selfish $B$ state; a smaller region leads to the altruistic $A$ state.
  }
  \label{fig:combined_vector_fields}
\end{figure}

Numerical analysis of Eq.~\eqref{eq:drhoD_uniqueu} in the $\rho_D = 0$ limit reveals either two or three stable equilibria depending on the infection cost $W_{\text{inf}}$ and the cleaning cost $W_{\text{cln}}$ (Fig.~\ref{fig:combined_vector_fields}). One stable equilibrium is always the pure altruistic state dominated by strategy $A$ ($\rho_A^* = 1$); the second one is the pure selfish state dominated by strategy B ($\rho_B^* = 1$).
The third equilibrium, which is only present for some parameter combinations, is a mixed state with a finite fraction of cheaters: $\rho_C ^* = 1-\rho_A^*-\rho_B^* > 0$.

When the infection cost is low, the mixed state is stable and in fact its basin of attraction can dominate evolutionary dynamics (Fig.~\ref{fig:combined_vector_fields}a). The picture remains qualitatively similar if the infection prevalence $x$ is decreased to $0.2$ (Fig.~\ref{fig:combined_vector_fieldsxdif}a) or increased to $0.8$ (Fig.~\ref{fig:combined_vector_fieldsxdif}b).
As the infection costs rise, the mixed state gets destabilized and its region of attraction gets subsumed by that of the selfish $B$ state (Fig.~\ref{fig:combined_vector_fields}b). There is a marked asymmetry between the large basin of attraction of the selfish state $B$ and the much smaller basin of attraction of the altruistic state $A$ in the high-infection-cost regime.
Note that while the vector flows vanish in the mixed stable state, they are non-zero in the purely altruistic and purely selfish states. These non-zero fluxes drive the population to the corresponding boundaries and are explicitly regularized by the boundary conditions, as described above.


We now relax the \(\rho_{D}=0\) assumption and study evolutionary dynamics in the vicinity of the purely altruistic state \(\rho_{A} = 1\): $\rho_A \approx 1;~\rho_B,\rho_C,\rho_D \ll 1$. The linearization of Eq.~\eqref{eq:drhoD_uniqueu} around the altruistic state yields
\begin{eqnarray} \label{eq:Acorner_FD_lin}
\dot{\rho}_A &=& \frac{1}{4} + \left(W_{\text{cln}}+\frac{1}{4}\right)(1-\rho_A) -
        \left(W_{\text{cln}}+\frac{3}{8}\right)\rho_B -
        \left(2W_{\text{cln}}+\frac{1}{2}\right)\rho_D + O(2), \\
\dot{\rho}_B &=& \frac{3}{8}(1-\rho_A) - \frac{3}{8}\rho_D + O(2), \nonumber \\
\dot{\rho}_D &=& \frac{1}{4} + \frac{1}{2}(1-\rho_A) - \frac{3}{8}\rho_B +
\left(W_{\text{cln}} - \frac{3}{4}\right)\rho_D + O(2), \nonumber
\end{eqnarray}
where \(O(2)\) denotes terms of second order in \((1-\rho_A,\rho_B,\rho_D)\). Note that we used Eq.~\eqref{eq:rho_sum} to eliminate $\rho_C$.
We find that the linearized dynamics are independent of the infection cost \(W_{\text{inf}}\). Indeed, despite a significant difference in the values of \(W_{\text{inf}}\) in Figs.~\ref{fig:combined_vector_fields}a and \ref{fig:combined_vector_fields}b, the green regions are similar.
Eq.~\eqref{eq:Acorner_FD_lin} shows that $\rho_{A} = 1$ is a stable fixed point: as $1-\rho_{A},\rho_{B},\rho_{D} \to 0$, $\dot{\rho}_{A} \to 1/4$ and thus remains positive.

A population in the purely altruistic state is vulnerable to an invasion by cheaters. Under the assumption $\rho_{B}=\rho_{D}=0$, we have $\rho_{C}=1-\rho_{A}$ and the first line in Eq.~\eqref{eq:Acorner_FD_lin} reduces to
\begin{equation} \label{eq:FA_edge_linear}
\dot{\rho}_A(\rho_C) \;=\; \frac{1}{4} \;+\; \Bigl(W_{\text{cln}}+\frac{1}{4}\Bigr)\rho_C \;+\; \mathcal{O}(\rho_C^2).
\end{equation}

In this limit, the population is no longer able to evolve toward the fully altruistic state when $\dot{\rho}_{A}=0$, which, according to Eq.~\eqref{eq:FA_edge_linear}, corresponds to
\begin{equation} \label{eq:rhoC_threshold_linear}
\;\rho_C^\ast \;\approx\; -\,\frac{1}{\,4W_{\text{cln}}+1\,}. 
\end{equation}
Since this approximation is valid in the $\rho_C^\ast\ll 1$ limit, it corresponds to large cleaning costs: $|W_\text{cln}| \gg 1$.
We note that $\rho_C^\ast$ in Eq.~\eqref{eq:rhoC_threshold_linear} is independent of $W_{\text{inf}}$. This result is in fact not limited to the linear approximation since Eqs.~\eqref{eq:drhoD_uniqueu} are independent of $W_{\text{inf}}$ if $\rho_{B}=\rho_{D}=0$.

The fraction of dutiful citizens converges to a small but finite value:
\begin{equation}
  \label{eq:3}
  \rho_{D} \approx \frac{1}{3-4W_{\text{cln}}}.
\end{equation}
This expression, obtained by setting $\dot{\rho}_{D} = 0$ in the last line of Eq.~\eqref{eq:Acorner_FD_lin}, is also valid in the $|W_\text{cln}| \gg 1$ limit.

We now linearize Eq.~\eqref{eq:drhoD_uniqueu} around the fully selfish state with $\rho_B = 1$:
\begin{eqnarray} \label{eq:Bcorner_FD_lin}
\dot{\rho}_A
&=& \frac{W_{\text{inf}}}{4}\,\rho_A
  + \frac{3}{8}\,(1-\rho_B)
  - \frac{3}{8}\,\rho_D
  + \mathcal{O}(2),
  \\
\dot{\rho}_B
&=& \frac{1}{4}
  - \frac{3}{8}\,\rho_A
  + \Bigl(W_{\text{cln}}-\frac{W_{\text{inf}}}{4}+\frac{1}{4}\Bigr)\,(1-\rho_B)
  + \Bigl(-\,2\,W_{\text{cln}}+\frac{W_{\text{inf}}}{4}-\frac{1}{2}\Bigr)\,\rho_D
  + \mathcal{O}(2),
  \nonumber \\
\dot{\rho}_D
&=& \frac{1}{4}
  - \frac{1}{4}\,\rho_A
  + \frac{3}{8}\,(1-\rho_B)
  + \Bigl(W_{\text{cln}}-\frac{5}{8}\Bigr)\,\rho_D
  + \mathcal{O}(2), \nonumber
\end{eqnarray}
Note that in the selfish-population limit, the population dynamics depend on both $W_{\text{inf}}$ and $W_{\text{cln}}$. Furthermore, $\rho_{B} = 1$ is a stable fixed point: as $1-\rho_{B},\rho_{A},\rho_{D} \to 0$, $\dot{\rho}_{A} \to 1/4$ and thus remains positive.

A selfish population is susceptible to an invasion by cheaters; the cheater-invasion threshold is given by
\begin{equation} \label{eq:rhoC_smallroot}
\rho_C^\ast
\;\approx\;
-\frac{1}{\,4W_{\text{cln}}-W_{\text{inf}}+1\,},
\end{equation}
which follows by setting $\dot{\rho}_B = 0$ in the second line of Eq.~\eqref{eq:Bcorner_FD_lin} and assuming that
$\rho_{A}=\rho_{D}=0$, such that $\rho_{C}=1-\rho_{B}$. Eq.~\eqref{eq:rhoC_smallroot} remains valid to leading order as long as
$\rho_{A},\rho_{D} \ll 1$. Since Eq.~\eqref{eq:Bcorner_FD_lin} and, consequently, Eq.~\eqref{eq:rhoC_smallroot} are valid in the
$\rho_C^\ast \ll 1$ limit, either $|W_\text{cln}|$ or $|W_\text{inf}|$ (or both) must be large.

The fraction of dutiful citizens, obtained by setting $\dot{\rho}_{D} = 0$ in the last line of Eq.~\eqref{eq:Bcorner_FD_lin}, converges to
\begin{equation} \label{eq:rhoD:xtra}
  \rho_{D} \approx \frac{2}{5 - 8W_{\text{cln}}},
\end{equation}
valid in the $|W_\text{cln}| \gg 1$ limit.

\begin{figure}[!htb]
  \centering
  \begin{minipage}[b]{0.48\textwidth}
    \begin{overpic}[width=\textwidth]{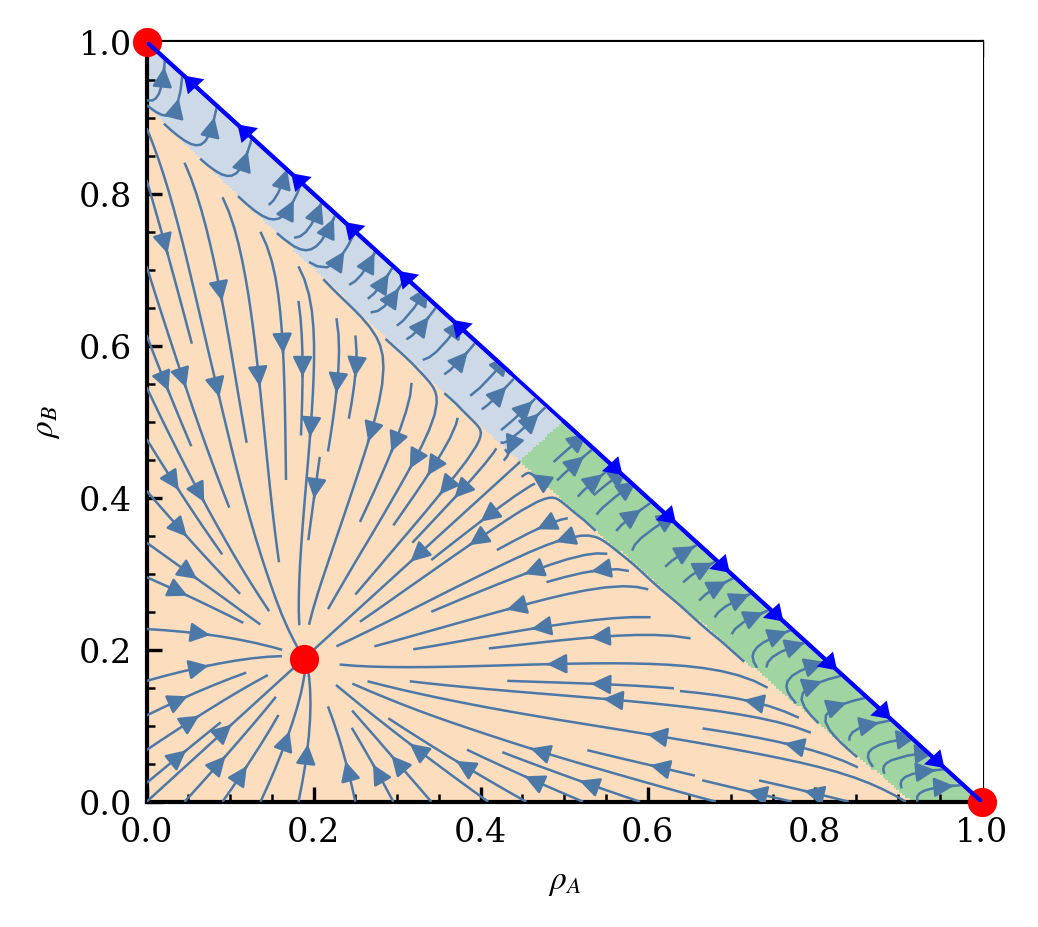}
      \put(4,94){\makebox(0,0)[l]{\footnotesize (a)}}   
    \end{overpic}
  \end{minipage}
  \hfill
  \begin{minipage}[b]{0.48\textwidth}
    \begin{overpic}[width=\textwidth]{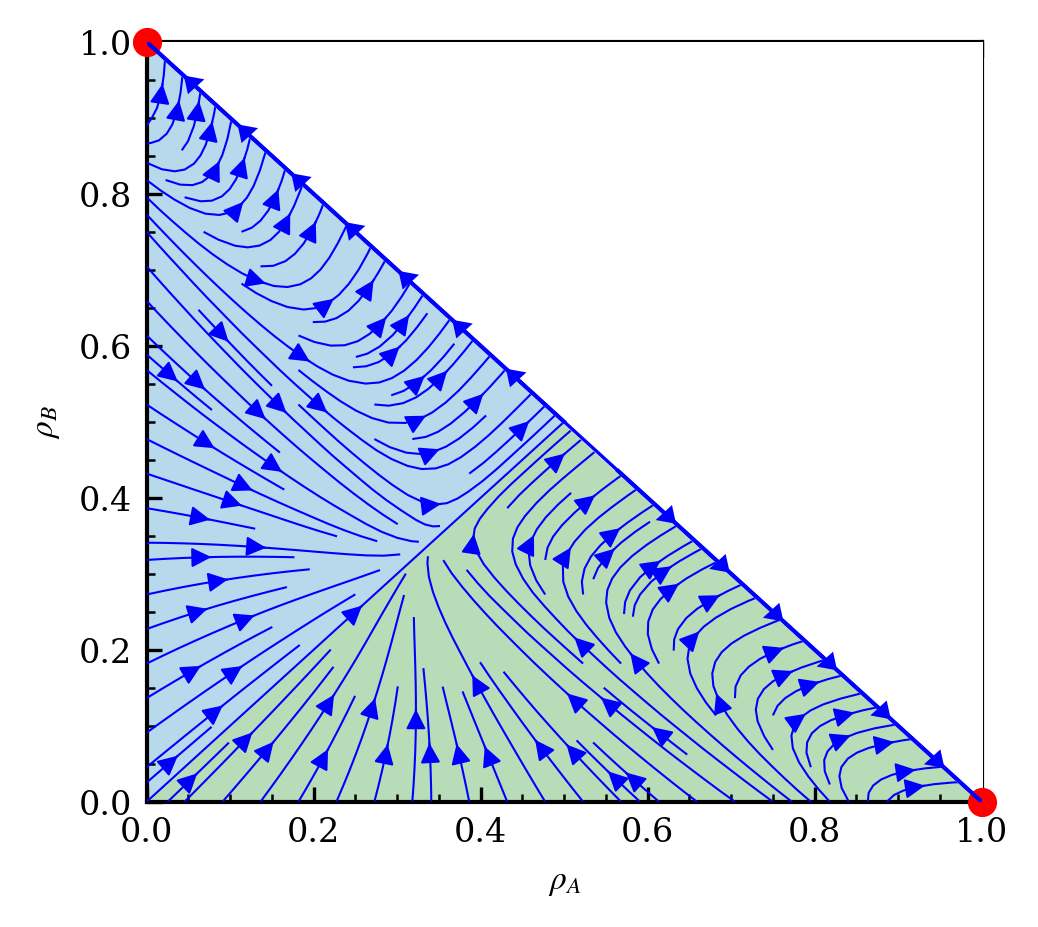}
      \put(4,94){\makebox(0,0)[l]{\footnotesize (b)}}   
    \end{overpic}
  \end{minipage}
\caption{ \textbf{Evolutionary dynamics and fixed point stability in the       absence of infection costs}. Altruistic ($A$)
    and selfish ($B$) strategies are symmetric when $W_{\text{inf}}=0$, since $f_A = f_B$ in this case (Eq.~\eqref{eq:fD_unique1}).
    As in Fig.~\ref{fig:combined_vector_fields}, we show vector fields over the strategy frequency space
    $(\rho_A,\rho_B)$, assuming that $\rho_{D}=0$ and hence $\rho_{C}=1-\rho_{A}-\rho_{B}$. Blue arrows depict the flows driven by selective forces and social pressure (Eq.~\eqref{eq:drhoD_uniqueu}). Red dots mark stable equilibria.
    (a) High cleaning cost: $W_{\text{cln}}=-4$.
    There are three stable equilibria: an altruistic monomorphic $A$ state ($\rho_{A}^*=1$), a selfish monomorphic
    $B$ state ($\rho_{B}^*=1$), and a mixed state comprising strategies
    $A$, $B$, and $C$. Background colors (green, blue, coral) denote numerically mapped regions of attraction of the altruistic, selfish,
    and mixed states, respectively. Most initial conditions converge to the mixed state. Note that the green and blue regions are equal in size.
    (b) Low cleaning cost: $W_{\text{cln}}=-2.5$.
    There are two stable equilibria: an altruistic monomorphic $A$ state ($\rho_{A}^*=1$) and a selfish monomorphic $B$ state ($\rho_{B}^*=1$).
    The mixed state becomes unstable in this regime and cheaters $C$ go extinct.
    Background colors (green, blue) denote numerically mapped regions of attraction of the altruistic and selfish states, respectively. Note that $A$ and $B$ regions of attraction are equal in size.
  }
\label{fig:dfdsf}
\end{figure}

Fig.~\ref{fig:dfdsf} shows evolutionary dynamics in the case of vanishing infection costs, $W_{\text{inf}} \to 0$. In this case, the cleaning costs control the number of stable fixed points. For higher cleaning costs, a mixed state with a finite fraction of cheaters is stable (Fig.~\ref{fig:dfdsf}a). For lower cleaning costs, the mixed state disappears and the population becomes either fully altruistic or fully selfish; the two basins of attraction are equal in size (Fig.~\ref{fig:dfdsf}b). The boundary between the mixed state and the $A$ and $B$ states can be approximated analytically under the assumption that it is a straight line (Appendix~\ref{sec:soci-barr-init}); the approximation closely matches the numerical estimate. The dynamical flows shown in Figs.~\ref{fig:dfdsf}a,b indicate hysteresis in the evolutionary dynamics: if the population is in the stable mixed state as in Fig.~\ref{fig:dfdsf}a and the costs change from high to low, destabilizing the mixed state, and then back again, the population does not necessarily return to the stable mixed state. Rather, it can remain in a fully altruistic or selfish state, which are both
characterized by finite basins of attraction.

\begin{figure}[t]
    \centering
    \begin{minipage}[b]{0.48\textwidth}
        \centering
        \begin{overpic}[width=\textwidth]{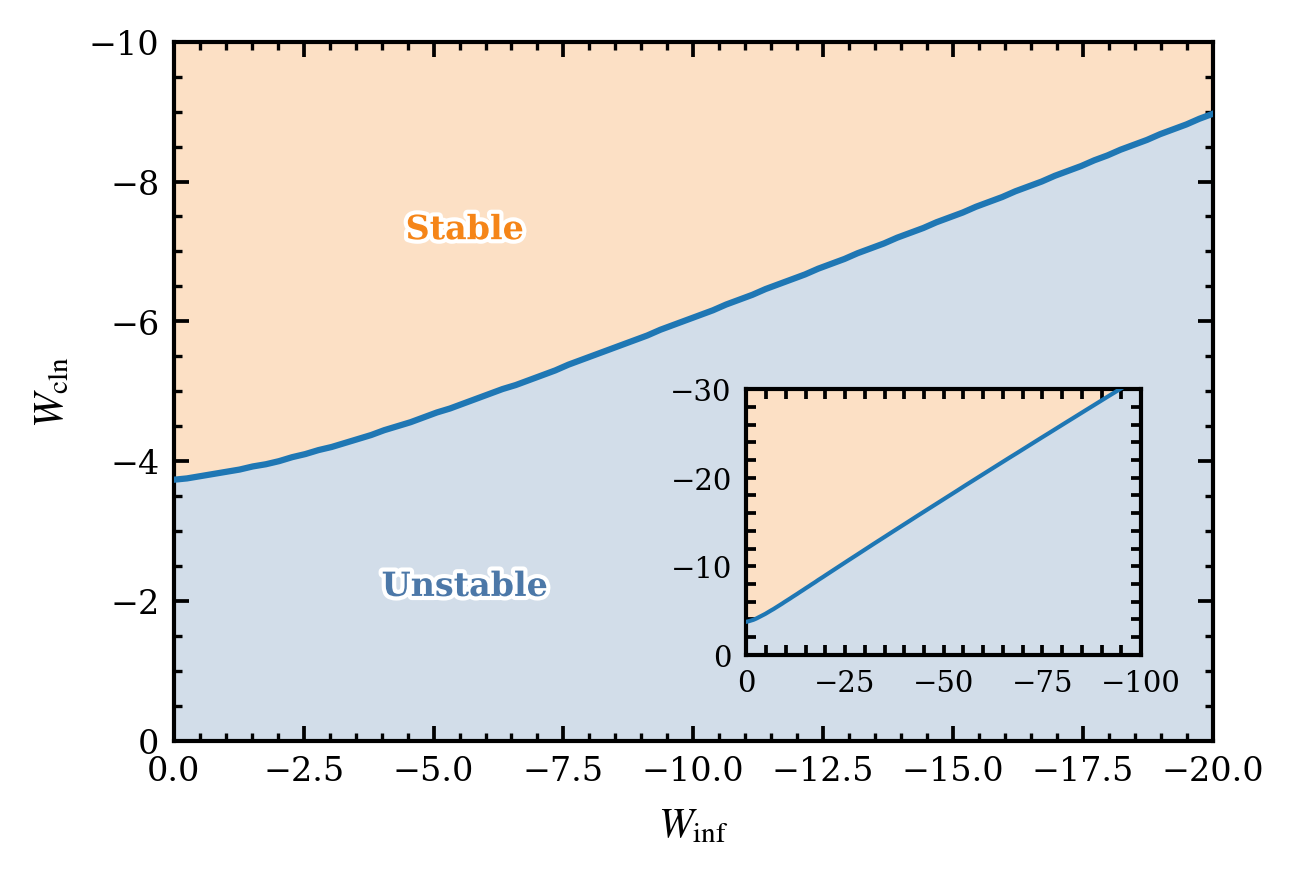}
           \put(4,70){\makebox(0,0)[l]{\footnotesize (a)}}   
        \end{overpic}
    \end{minipage}
    \hfill
    \begin{minipage}[b]{0.48\textwidth}
        \centering
        \begin{overpic}[width=\textwidth]{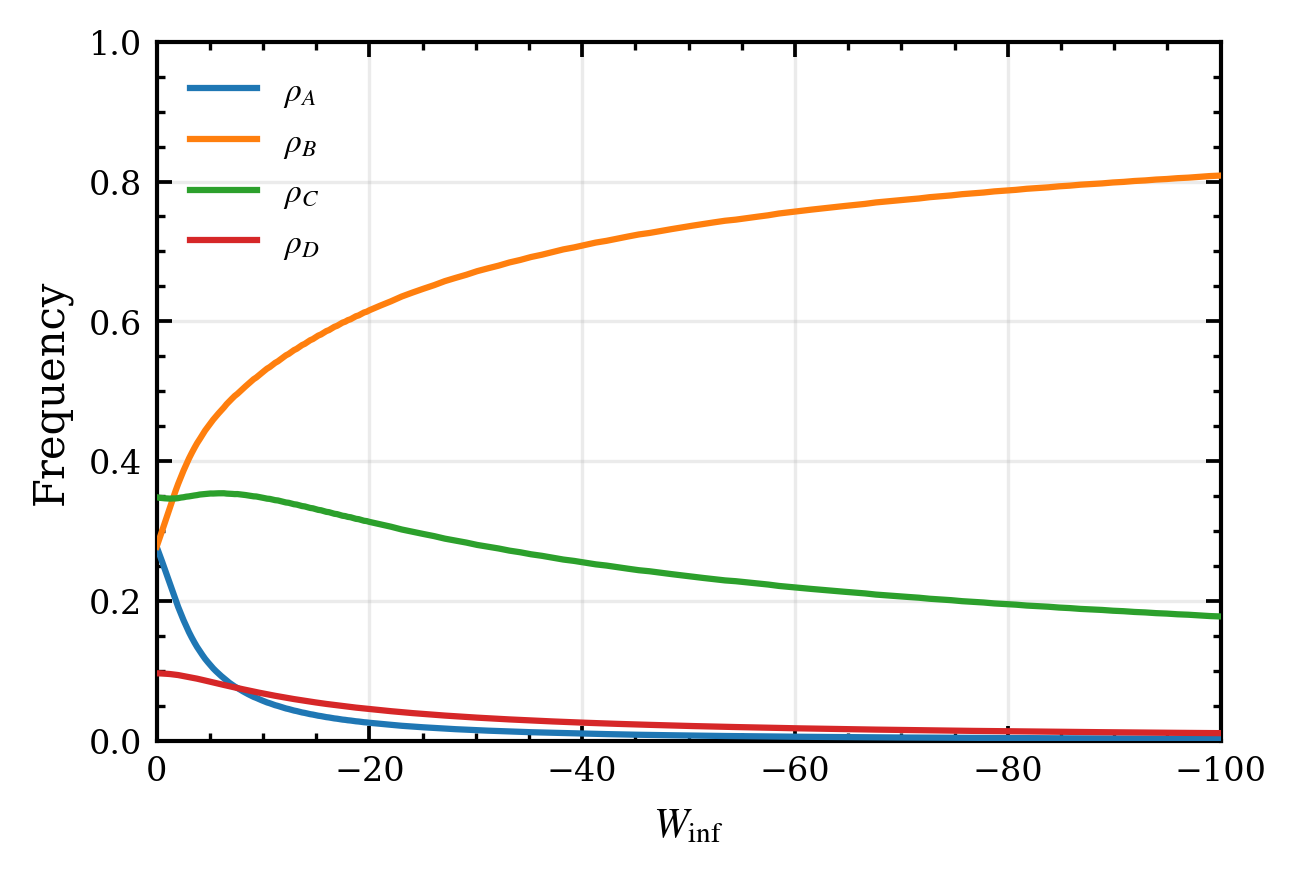}
          \put(4,70){\makebox(0,0)[l]{\footnotesize (b)}}   
        \end{overpic}
    \end{minipage}
\caption{ \textbf{Mixed-state stability and composition.}
  Here we allow all four strategies to have non-zero frequencies.
  (a) Stability of the mixed state in the cost parameter space $(W_{\text{inf}},W_{\text{cln}})$. The blue region indicates an unstable mixed state, where cheaters go extinct irrespective of initial conditions. The coral region marks cost combinations that permit coexistence of cheaters with other strategies. The insert shows a zoomed-out view in the larger $(W_{\text{inf}},W_{\text{cln}})$ domain.
  (b) Strategy frequencies as functions of $W_{\text{inf}}$ along the stability boundary in the inset of panel (a). At each value of $W_{\text{inf}}$, the stability boundary corresponds to the minimum fraction of cheaters allowed in a mixed state. As the boundary is crossed, the fraction of cheaters drops abruptly to zero.
}
\label{fig:combined_dynamics}
\end{figure}

Fig.~\ref{fig:combined_dynamics}a shows the stability of the mixed state as a function of the infection cost $W_{\text{inf}}$ and the cleaning cost $W_{\text{cln}}$ in the more general case with $\rho_D \ne 0$. We find computationally that the number and the nature of fixed points remain the same as in the $\rho_D = 0$ case, consistent with the assumption that $D$-type individuals play a minor role in population dynamics. As in the $\rho_D = 0$ case illustrated in Figs.~\ref{fig:combined_vector_fields} and \ref{fig:dfdsf}, there is a region with a single stable fixed point corresponding to the mixed state. This fixed point becomes unstable with decreasing $|W_{\text{cln}}|$ and/or increasing $|W_{\text{inf}}|$, leading to the abrupt disappearance of cheaters from the population. The frequencies at the stability boundary of the mixed state show that altruists and dutiful citizens quickly disappear from the population as infection costs increase (Fig.~\ref{fig:combined_dynamics}b). Above $|W_{\text{inf}}| > 40$, the population is essentially divided between cheaters and selfish individuals, with the fraction of cheaters gradually decreasing with $|W_{\text{inf}}|$. Asymptotically, $\rho_{B} \to 1$ as $W_{\text{inf}} \to -\infty$. Thus, the population at the boundary can transition smoothly from a mixed state to the purely selfish state by increasing the costs of infection.
Note that the stable mixed state with $W_{\text{cln}}=-4$, $W_{\text{inf}}=-5$ in Fig.~\ref{fig:combined_vector_fields}a becomes unstable when $\rho_D$ is allowed to take on non-zero values, indicating that the $\rho_D = 0$ assumption leads to a moderate shift in the boundary between the two regions.



\begin{figure}[t]
  \centering
  \begin{minipage}[b]{0.48\textwidth}
    \centering
    \begin{overpic}[width=\textwidth]{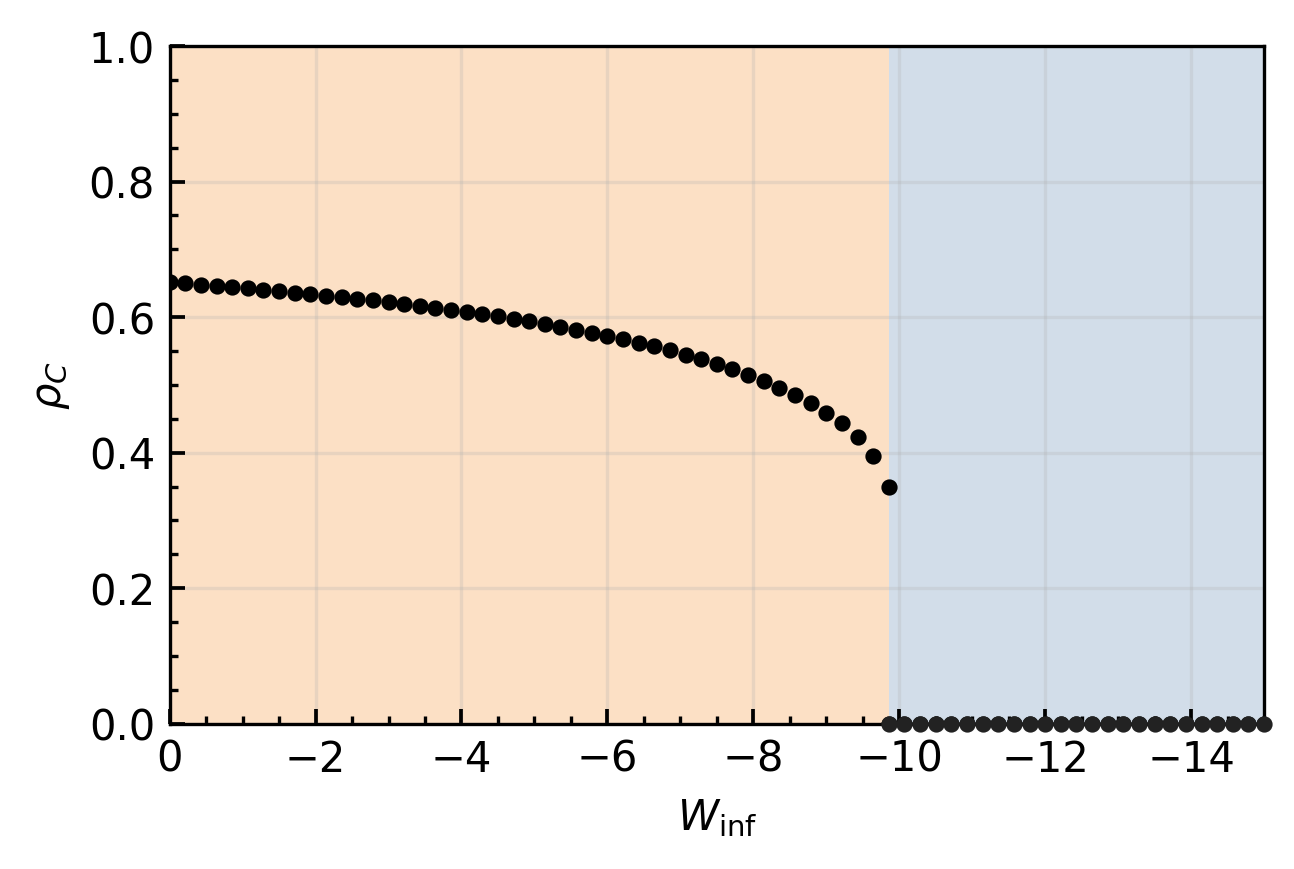}
            \put(4,70){\makebox(0,0)[l]{\footnotesize (a)}}   
    \end{overpic}
  \end{minipage}
  \hfill
  \begin{minipage}[b]{0.48\textwidth}
    \centering
    \begin{overpic}[width=\textwidth]{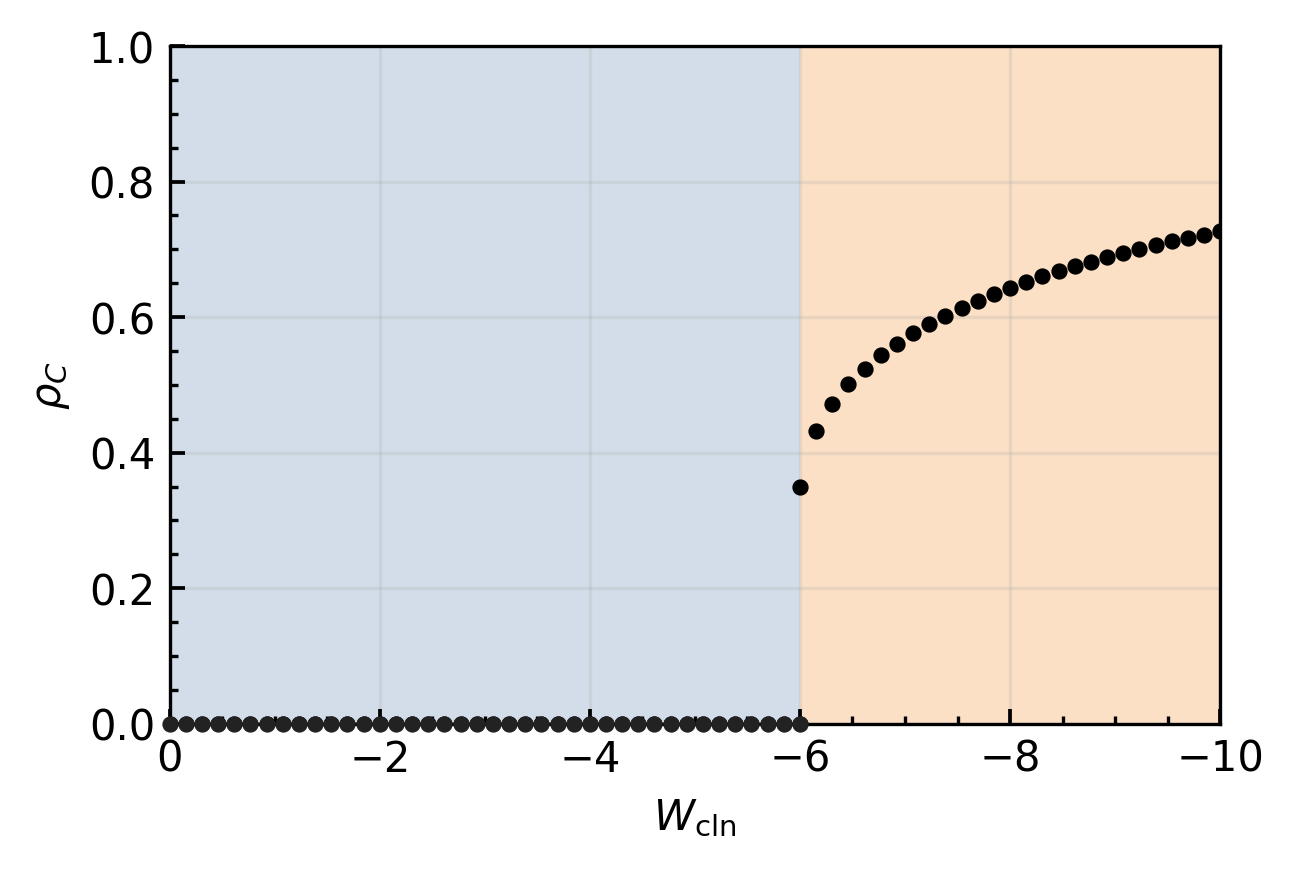}
            \put(4,70){\makebox(0,0)[l]{\footnotesize (b)}}   
    \end{overpic}
  \end{minipage}
\caption{ \textbf{Abrupt elimination of cheaters as a function of infection and cleaning costs.} We observe an abrupt, phase-like transition between non-zero and zero cheater frequencies as the boundary between stable and unstable regions in Fig.~\ref{fig:combined_dynamics} is crossed. (a) The fraction of cheaters in the population, $\rho_C$, as a function of the infection cost $W_{\text{inf}}$, with the cleaning cost fixed at $W_{\text{cln}}=-6$. Starting from the initial condition $\rho_{A}^0=\rho_{B}^0=\rho_{D}^0=0.1$, we iterate the dynamics either to a fixed point 
or until cheaters are eliminated and plot the resulting $\rho_{C}$ values (black dots). A sharp transition occurs at $W_{\text{inf}}=-9.8$, matching the boundary between stable and unstable regions in Fig.~\ref{fig:combined_dynamics}; background colors indicate those regions. (b) The fraction of cheaters in the population, $\rho_C$, as a function of the cleaning cost $W_{\text{cln}}$, with the infection cost fixed at $W_{\text{inf}}=-9.8$. The values of $\rho_C$ for each $W_{\text{cln}}$ are obtained as in (a), using the same initial conditions. 
A sharp transition occurs at $W_{\text{cln}}=-6$.
  }
\label{fig:jhkjh}
\end{figure}

Cheaters are eliminated abruptly from mixed populations with $\rho_{C}>0$ as the costs of infection ($W_{\text{inf}}$) or cleaning ($W_{\text{cln}}$) vary gradually, for instance due to increased access to disinfectants, emergence of new virus variants, or immunity buildup in the population (Fig.~\ref{fig:jhkjh}). The phase-like transition occurs when the system crosses the stability boundary in Fig.~\ref{fig:combined_dynamics}a. 
For example, if the boundary is crossed horizontally (going right) with $W_{\text{cln}}=-6$ (Fig.~\ref{fig:jhkjh}a), a sharp transition to $\rho_C = 0$ occurs at $W_{\text{inf}}=-9.8$, in agreement with Fig.~\ref{fig:combined_dynamics}a. The frequencies at the stability boundary of the mixed state determine the minimum fraction of cheaters that can sustain the mixed state.
Likewise, if the boundary is crossed vertically (going up) with $W_{\text{inf}}=-9.8$ (Fig.~\ref{fig:jhkjh}b), a sharp transition to $\rho_C > 0$ occurs at $W_{\text{inf}}=-6$. These outcomes agree with the fixed-point stability analysis described above. The results in Figs.~\ref{fig:jhkjh}a,b depend on the initial conditions, here set to $\rho_{A}^0=\rho_{B}^0=\rho_{D}^0=0.1$ (yielding $\rho_{C}^0=0.7$). For example, if the population
starts near the altruistic state $\rho_{A}\approx 1$, a fluctuation in
$\rho_{C}$ exceeding the threshold value $\rho_C^\ast$ (Eq.~\eqref{eq:rhoC_threshold_linear}) is required to
destabilize it.


Together, our results indicate that simple, empirically motivated
assumptions can yield non-trivial predictions:
steady-state populations may be fully altruistic,
fully selfish, or contain a sizable fraction of cheaters.
Changes in infection and cleaning costs can trigger abrupt
transitions between qualitatively distinct states.
Our game-theoretic evolutionary model can be used to explain social behavior and inform policies for managing the hygiene of shared resources.

\section{Discussion and Conclusion}

This study has introduced an empirically grounded, interpretable model of collective hygiene in shared-use environments. Our model uses a realistic set of strategies available to all individuals in their equipment usage: clean after ($A$), clean before ($B$), neither ($C$), or both ($D$). We find that for a broad range of initial conditions and cleaning and infection costs, the population reaches a steady state with a sizable share of free riders who employ strategy $C$ (i.e., cheaters who do not clean equipment at all). However, this `mixed' state can be destabilized when the costs are varied, producing an abrupt, phase-like transition to a cheater-free steady state (Figs.~\ref{fig:combined_dynamics}a,~\ref{fig:jhkjh}). The transitions can be path-dependent and exhibit such non-linear phenomena as lock-in and hysteresis~\cite{Strogatz:2024}. If the mixed state is stable, it can be reached from cheater-free states only if a critical fraction of cheaters is introduced into the population; subcritical perturbations will be removed by evolutionary dynamics and one of the cheater-free steady states established again.

In the altruistic steady-state in which all users practice strategy $A$, the critical fraction of cheaters $C$ required to destabilize the population is independent of the infection cost. This is a consequence of the shared infection pathway: cheaters and altruists alike acquire infection from prior users of shared equipment and are therefore affected similarly by the infection cost. This observation may have significant implications for designing public-health policies.

Within the mixed state, the selfish strategy $B$ generally dominates, and this dominance strengthens as the infection costs increase (Fig.~\ref{fig:combined_dynamics}b). The mixed state becomes unstable when infection costs are high (imposing a large penalty on cheaters) or cleaning costs are low (eliminating their advantage).
When infection costs are significant, the basin of attraction of the selfish $B$ state is typically greater than than of the altruistic $A$ state (Fig.~\ref{fig:combined_vector_fields}). However, in the absence of infection costs the system becomes symmetric
(Fig.~\ref{fig:dfdsf}). In the case of high cleaning and low infection costs, the basin of attraction of the mixed state predominates those of the cheater-free states (Fig.~\ref{fig:dfdsf}a).


Our model is well-suited to empirical validation through behavioral analysis, including quantitative observation of cleaning practices under experimentally controlled conditions. In particular, our observations are in a qualitative agreement with a recent study which examined post-use cleaning practices, reporting significant increases in post-cleaning levels when prompts are provided and access to cleaning materials is improved~\cite{Elba2018}. These findings broadly align with our conclusions: at any fixed infection cost, cheaters are gradually and then abruptly removed from the population as the cleaning costs decrease in magnitude, resulting in the overall increase in the use of cleaning products (Figs.~\ref{fig:combined_dynamics}a and \ref{fig:jhkjh}b). For low infection costs, the model predicts an expansion of the basin of attraction for the post-use cleaning state $A$ (cf. green regions in Figs.~\ref{fig:dfdsf}a and \ref{fig:dfdsf}b).
In both fully altruistic and fully selfish cases, monitoring the share of cheaters relative to the model's invasion thresholds is crucial; crossing these thresholds risks a sudden regime shift.

The model presented here is, to the best of our knowledge, the first quantitative description of infection maintenance and propagation in shared-use environments. Future work can incorporate various refinements and extensions into our basic framework: (a) spatial structure of user populations~\cite{Nowak2006a,Ohtsuki2006}; (b) heterogeneity of risk perception; (c) reciprocity and reputation~\cite{Trivers1971,Nowak2006a,Hauser2009,Schmid2021}; (d) consequences of stringent interventions such as quarantines and shutdowns. It would also be interesting to couple our model with an explicit treatment of epidemiological dynamics, e.g. based on SIR-type models~\cite{Lopez2025}. Furthermore, it is possible to incorporate explicit decision-making and alternative evolutionary dynamics that would replace replicator dynamics, such as Brown--von Neuman--Nash~\cite{brown1951fictitious,gilboa1991social,Sandholm2010a}, Smith~\cite{Smith1982,Sandholm2010a}, or logit dynamics~\cite{fudenberg1998tlg,Sandholm2010a}. 
Finally, there has been significant recent focus on the environmental feedback that modifies payoffs in replicator dynamics~\cite{Weitz2016,Tilman2020,Wang2020,Chen2018}. Such payoffs can be potentially integrated into our model, which currently employs constant infection and cleaning costs augmented by a gradient term that represents societal pressure as a potential game~\cite{Monderer1996,Friedman2010,Taylor1997}.

In conclusion, our work has extended the analysis of emergence and stability of social cooperation~\cite{Axelrod1981,Hofbauer1998a} to sequential interactions involving shared resources. We have characterizes the conditions under which altruistic, selfish, and mixed hygiene regimes arise; explained the mechanisms that drive abrupt regime shifts; and identified conditions for crossing stability thresholds. If we treat ``infection'' as any propagating risk and ``cleaning'' as maintenance, our framework can be applied not only to gyms but to vehicle fleets, co-working spaces, hotels, and electronic devices -- and, by analogy, to civic and national well-being, offering potential insights into why some communities prosper while others fail~\cite{Acemoglu2012}.

\section*{Acknowledgments}
A.V.M. acknowledges financial and logistical support from the Center for Quantitative Biology, Rutgers University. A.F. recognizes the Interdisciplinary Computational Physics Laboratory at the Racah Institute of Physics for the allocation of computational resources and technical assistance.


\appendix
\section{Derivation of infection probabilities.} \label{sec:deriv_indiv_prob}

\textbf{The social pressure potential.}
Recall that infection probabilities apply only to individuals employing strategies $A$ and $C$; individuals adhering to strategies $B$ and $D$ are protected because they clean the equipment before using it. 
To compute the total number of infections among $C$-type users, we note that a $C$-type user can be infected only if they are preceded by an infected user of type $B$ or $C$. Users of types $A$ and $D$ clean the equipment after use and therefore protect the following user. Both the total number of infections and the infection probabilities depend on the number of pieces of equipment, $N_{p}$, and on the number of successive uses of a single piece of equipment within a given time interval, $N_{u}$. We assume that each individual uses a piece of equipment once before moving on. Moreover, we assume that all pieces of equipment are in continuous use and that users have no preference for one type of equipment over another.


Consider contamination events affecting type-$C$ users on a single shared device used sequentially $N_{u}$ times. A contamination chain begins when an infected type-$B$ or type-$C$ user uses the device, continues through consecutive use by $K$ uninfected type-$C$ users, and ends when the next user is anyone other than an uninfected type-$C$ user. Starting from a clean device, such a chain infects $K$ type-$C$ users. Summing over all possible infection chains, the total number of infection events among type-$C$ users is given by:
\begin{equation}
M^{tot}_C = N_{p} N_{u} \sum_{K=0}^{\infty} x(\rho_B + \rho_C) \rho_C^K (1 - x)^K
        \left[1 - \rho_C (1 - x)\right] K,
        \label{eq:MC_sum_unique}
\end{equation}
where $x(\rho_B + \rho_C)$ is the probability that the chain is initiated by an infected type-$B$ or type-$C$ individual, $\rho_C^K (1-x)^K$ is the probability that the chain has length $K$ consisting of uninfected type-$C$ users, and $\left[1 - \rho_C (1 - x)\right]$ is the probability that the next user is not an uninfected type-$C$ user. As previously discussed, we assume that contact with contaminated equipment transmits infection with probability one.

Note that the factor of $K$ is necessary in Eq.~\eqref{eq:MC_sum_unique} because we count each contamination chain once starting from its beginning; each chain entails $K$ contamination events whose probabilities need to be added up. 
Furthermore, extending the upper limit of the sum to
infinity on the right-hand side of Eq.~\eqref{eq:MC_sum_unique}
requires that the mean of $K$ be much smaller than the total number of
uses of a single piece of equipment, $\langle K \rangle  =  \sum_{K=0}^{\infty} \rho_C^K (1 - x)^K
        \left[1 - \rho_C (1 - x)\right] K \ll N_{u}$.
        

Similarly to Eq.~\eqref{eq:MC_sum_unique}, the total number of infections in $A$-type individuals is given by
\begin{equation} \label{eq:MA_sum_unique}
M^{tot}_A = N_{p} N_{u} \sum_{K=0}^{\infty} x(\rho_B + \rho_C) \rho_C^K (1 - x)^K \rho_A (1 - x).
\end{equation}

Using the series identities
\begin{equation} \label{eq:series_identities_unique}
\sum_{k=0}^{\infty} k y^k = \frac{y}{(1 - y)^2}; \quad \sum_{k=0}^{\infty} y^k = \frac{1}{1 - y}, 
\end{equation}
the expressions for $M_A^{tot}$ and $M_C^{tot}$ simplify to
\begin{equation} \label{eq:MA_final_unique}
M_A^{tot} = N_{p} N_{u} x(1 - x) \rho_A (\rho_B + \rho_C) \frac{1}{1 - \rho_C (1 - x)}
\end{equation}
and
\begin{equation} \label{eq:MC_final_unique}
M_C^{tot} = N_{p} N_{u} x(1 - x)(\rho_B + \rho_C) \frac{\rho_C}{1 - \rho_C(1 - x)}.
\end{equation}

The infection probability of a type-$A$ individual after a single
instance of equipment usage is then given by: 
\begin{equation}
I_A = \frac{M_A^{tot}}{N_{p} N_{u} \rho_A} =  \,x(1 - x)(\rho_B +
\rho_C)\,\frac{1}{1 - \rho_C (1 - x)}, \label{eq:IA_unique5}
\end{equation}
where $N_{u} \rho_{A}$ is the total number of type-$A$ individuals who used the equipment over a given interval of time.
Similarly, for a type-$C$ individual the infection probability after a single instance of equipment usage is given by:
\begin{equation}
I_C = \frac{M_C^{tot}}{N_{p} N_{u} \rho_C} = \, x(1 - x)(\rho_B + \rho_C)\,\frac{1}{1 - \rho_C (1 - x)}. \label{eq:IC_final_unique}
\end{equation}
Note that the two infection probabilities are identical, $I_C = I_A = I$, because the infection pathways are the same in both cases.
Straightforward calculations show that $0 \le I \le 1$.

Finally, we define the social pressure potential as:
\begin{equation}
\label{eq:TotalRate}
\Phi_{\text{soc}}  = \frac{1}{\sigma} \left( I_A \rho_A + I_C \rho_C \right) = \frac{1}{\sigma}\frac{(\rho_B + \rho_C)(\rho_{A}+\rho_{C})x(1 - x)}{1 - \rho_C(1 - x)},
\end{equation}
where $\sigma$ sets the overall strength of social pressure.




\textbf{The role of infection prevalence $x$.}
Throughout most of this work, the infection prevalence in the population was set to $x = 0.5$, corresponding to the maximum in the $x(1-x)$ prefactor in Eq.~\eqref{eq:I_rhoB_rhoC_x}. We find that the specific choice of $x$ affects our results only moderately and does not change the main conclusions of the article (Fig.~\ref{fig:combined_vector_fieldsxdif}). 


Indeed, let us rescale several variables in Eq.~\eqref{eq:drhoD_uniqueu} by $f(x) = x(1-x)$: $I^\ast = I/f(x)$, $W_{\text{cln}}^{\ast}=W_{\text{cln}}/f(x)$,
$\Phi_{\text{soc}}^\ast = \Phi_{\text{soc}}/f(x)$, $t^{\ast}=t f(x)$. Then Eq.~\eqref{eq:drhoD_uniqueu} becomes:
\begin{eqnarray} \label{eq:drhoD_uniqueuapp} 
\dot{\rho}_A &=& \rho_A (f_A (W_\text{cln}^\ast, W_\text{inf}, I^\ast) -
                 \bar{f}) - \sigma \frac{\partial \Phi_{\text{soc}}^\ast}{\partial \rho_{A}},
                   \\
\dot{\rho}_B &=& \rho_B (f_B (W_\text{cln}^\ast) -
                 \bar{f}) - \sigma \frac{\partial \Phi_{\text{soc}}^\ast}{\partial \rho_{B}}, \nonumber \\
\dot{\rho}_C &=& \rho_C (f_C (W_\text{inf}, I^\ast) -
                 \bar{f}) - \sigma \frac{\partial \Phi_{\text{soc}}^\ast}{\partial \rho_{C}}, \nonumber \\
\dot{\rho}_D &=& \rho_D (f_D (W_\text{cln}^\ast) -
\bar{f}) - \sigma \frac{\partial \Phi_{\text{soc}}^\ast}{\partial \rho_{D}}, \nonumber
\end{eqnarray}
where the time derivatives are now with respect to $t^{\ast}$. Eq.~\eqref{eq:drhoD_uniqueuapp} is almost the same as the rescaled Eq.~\eqref{eq:drhoD_uniqueu} with $x=0.5$ -- the only difference is that $1 - \rho_C/2$ is replaced by $1 - \rho_C (1 - x)$ in the denominator of the social potential (cf. Eq.~\eqref{eq:TotalRate}). This difference is negligible when $\rho_C \ll 1$, and in general its influence is limited because $1-x$ only changes in the $[0,1]$ range. To summarize, Eq.~\eqref{eq:drhoD_uniqueuapp} provides a starting point for quantitative analysis of systems with $x \ne 0.5$. However, we expect the deviations from the $x=0.5$ case explored in detail in this work to be relatively minor.



\section{Evolutionary dynamics in the absence of infection costs.} \label{sec:soci-barr-init}

To determine the boundary of the basin of stability of the mixed state, we note that the vector flows are approximately tangent to the line
$\rho_{A} + \rho_{B} = \text{const}$ at the mixed-state boundary in Figs.~\ref{fig:combined_vector_fields}a and~\ref{fig:combined_vector_fieldsxdif}. Thus,
\begin{equation}
  \dot{\rho}_{A} = -\dot{\rho}_{B},
\label{eq:6}
\end{equation}
at the boundary.
Eq.~\eqref{eq:6} reduces to a fourth-order polynomial in $\rho_{A}$ and $\rho_{B}$ (cf. Eq.~\eqref{eq:drhoD_uniqueu}; note also that $\rho_{C}^{*} = 1 - \rho_A - \rho_B$, where $\rho_{C}^{*}$ is the approximately constant fraction of cheaters at the stability boundary):
\begin{eqnarray} \label{eq:11}
&-&2 \rho_A^{4} W_\text{cln}
- 2 \rho_A^{3} W_\text{cln}
+ \rho_B^{2}\bigl(
    -2 \rho_A^{2} W_\text{inf}
    - 12 \rho_A^{2} W_\text{cln}
    + 2 \rho_A W_\text{inf}
    - 6 \rho_A W_\text{cln}
    + 2 W_\text{cln}
    - 1
  \bigr) \nonumber \\
&+& 2 \rho_A^{2} W_\text{cln}
- \rho_A^{2}
+ \rho_B\bigl(
    - \rho_A^{3} W_\text{inf}
    - 8 \rho_A^{3} W_\text{cln}
    + \rho_A^{2} W_\text{inf}
    - 6 \rho_A^{2} W_\text{cln}
    + \rho_A W_\text{inf}
    + 4 \rho_A W_\text{cln}
    - W_\text{inf}
    + 2 W_\text{cln}
    - 1
  \bigr) \nonumber \\
&+& \rho_B^{3}\bigl(
    - \rho_A W_\text{inf}
    - 8 \rho_A W_\text{cln}
    + W_\text{inf}
    - 2 W_\text{cln}
  \bigr)
+ 2 \rho_A W_\text{cln}
- \rho_A
- 2 \rho_B^{4} W_\text{cln}
+ 4=0.
\end{eqnarray}
It therefore admits an analytical (though cumbersome) solution, which we use to plot the boundary of the basin of attraction of the mixed state.
Below we consider a few limiting cases which help clarify the general solution of Eq.~\eqref{eq:11}.

In the case $\rho_{A}=0$ and substituting $\rho_{B}=1-\rho_{C}^{*}$,
Eq.~\eqref{eq:11} becomes:
\begin{equation} \label{eq:22a}
-2 W_\text{cln}\,\rho_C^{*\,4}
+ \bigl(10 W_\text{cln} - W_\text{inf}\bigr)\,\rho_C^{*\,3}
- \bigl(16 W_\text{cln} - 3 W_\text{inf} + 1\bigr)\,\rho_C^{*\,2}
+ \bigl(8 W_\text{cln} - 2 W_\text{inf} + 3\bigr)\,\rho_C^{*}
+ 2 = 0.
\end{equation}

In the $\rho_C^{*} \ll 1$ limit (or, alternatively, when $|W_\text{cln}|$, $|W_\text{inf}|$, or both are large), we obtain:
\begin{equation} \label{eq:8}
   \rho_C^{*} \approx -\frac{1}{4 W_\text{cln} -  W_\text{inf}},
\end{equation}
consistent with Eq.~\eqref{eq:rhoC_smallroot}.

In the case $\rho_{B}=0$ and substituting $\rho_{A}=1-\rho_{C}^{\ast}$, Eq.~(\ref{eq:11}) becomes:
\begin{equation} \label{eq:22b}
(-2 W_\text{cln})\,\rho_C^{*\,4}
+ (10 W_\text{cln})\,\rho_C^{*\,3}
- (16 W_\text{cln} + 1)\,\rho_C^{*\,2}
+ (8 W_\text{cln} + 3)\,\rho_C^{*}
+ 2 = 0.
\end{equation}
In this case, $\rho_C^{*}$ no longer depends on the infection costs.
In the $\rho_C^{*} \ll 1$ (or, alternatively, $|W_\text{cln}| \gg 1$) limit, we obtain:
\begin{equation} \label{eq:9}
  \rho_C^{*} \approx -\frac{1}{4W_\text{cln}},
\end{equation}
consistent with Eq.~\eqref{eq:rhoC_threshold_linear}.

Finally, in the absence of infection costs (\(W_\text{inf}=0\)) we substitute \(\rho_{B}=1-\rho_{A}-\rho_{C}^{*}\) into Eq.~\eqref{eq:11}, which results in:
\begin{equation}
-2 W_\text{cln}\,\rho_{C}^{*\,4}
+ 10 W_\text{cln}\,\rho_{C}^{*\,3}
- (16 W_\text{cln} + 1)\,\rho_{C}^{*\,2}
+ (8 W_\text{cln} - 2\rho_A + 3)\,\rho_{C}^{*}
- 2\rho_A^{2} + 2\rho_A + 2
= 0.
\end{equation}
An asymptotic solution, valid in the $\rho_{C}^{*} \ll 1$, $|W_\text{cln}| \gg \rho_A$ limit, is given by:
\begin{equation} \label{eq:10}
\rho_{C}^{*} \approx \frac{\rho_A^{2} - \rho_A - 1}{4\,W_\text{cln}},
\end{equation}
consistent with Eqs.~\eqref{eq:8} and \eqref{eq:9}. When the cleaning costs are large, $|W_\text{cln}| \gg 1$, the curvature of the quadratic expression on the right-hand side of Eq.~\eqref{eq:10} is low and it is well approximated by a straight line, as assumed in Eq.~\eqref{eq:6}. In other words, \(\rho_{C}^{*}\) is nearly constant along the boundary in this case.

\bibliographystyle{apsrev4-2}
\bibliography{gym_refs}

@article{Nguyen2025,
  title = {Where Physics Meets Behavior in Animal Groups},
  author = {Nguyen, Chantal and Peleg, Orit},
  year = 2025,
  month = jul,
  journal = {PRX Life},
  volume = {3},
  number = {3},
  pages = {037001}
}

@article{Castellano2009,
  title = {Statistical Physics of Social Dynamics},
  author = {Castellano, Claudio and Fortunato, Santo and Loreto, Vittorio},
  year = 2009,
  month = may,
  journal = {Rev. Mod. Phys.},
  volume = {81},
  number = {2},
  pages = {591--646}
}

@article{Szabo2007,
  title = {Evolutionary Games on Graphs},
  author = {Szab{\'o}, Gy{\"o}rgy and F{\'a}th, G{\'a}bor},
  year = 2007,
  journal = {Phys. Rep.},
  volume = {446},
  pages = {97--216}
}

@article{Baer1968,
  title = {Some current dimensions of applied behavior analysis},
  author = {Donald M. Baer and Montrose M. Wolf and Todd R. Risley},
  journal = {J. Appl. Behav. Anal.},
  volume = {1},
  number = {1},
  pages = {91--97},
  year = {1968}
}

@book{Sandholm2010a,
  title = {Population Games and Evolutionary Dynamics},
  author = {Sandholm, William H.},
  year = {2010},
  publisher = {MIT Press},
  address = {Cambridge, MA}
}

@book{Hofbauer1998a,
  title = {Evolutionary Games and Population Dynamics},
  author = {Hofbauer, Josef and Sigmund, Karl},
  year = 1998,
  month = may,
  publisher = {Cambridge University Press},
  address = {Cambridge, UK}
}

@article{Axelrod1981,
  title = {The Evolution of Cooperation},
  author = {Robert Axelrod and W D Hamilton},
  journal = {Science},
  volume = {211},
  number = {4489},
  pages = {1390--1396},
  year = {1981}
}

@article{Nowak2006a,
  title = {Five Rules for the Evolution of Cooperation},
  author = {Nowak, Martin A.},
  year = 2006,
  month = dec,
  journal = {Science},
  volume = {314},
  number = {5805},
  pages = {1560--1563}
}

@article{Elba2018,
  title = {Increasing the Post-Use Cleaning of Gym Equipment Using Prompts and Increased Access to Cleaning Materials},
  author = {Elba, Ilexis and Ivy, Jonathan W.},
  year = 2018,
  month = dec,
  journal = {Behav. Anal. Pract.},
  volume = {11},
  number = {4},
  pages = {390--394}
}

@article{Austin1993,
  title = {Increasing recycling in office environments: {T}he effects of specific, informative cues},
  author = {John Austin and David B. Hatfield and Angelica C. Grindle and Jon S. Bailey},
  journal = {J. Appl. Behav. Anal.},
  volume = {26},
  number = {2},
  pages = {247--253},
  year = {1993}
}

@article{Fournier2012,
  title = {Effects of Response Cost and Socially-Assisted Interventions on Hand-Hygiene Behavior of University Students},
  author = {Angela K. Fournier and Thomas D. Berry},
  journal = {Behav. Soc. Issues},
  volume = {21},
  number = {1},
  pages = {152--164},
  year = {2012}
}

@article{Shawler2021,
  title = {A Proposed Functional Analysis of Transmission Prevention Behaviors for a Respiratory Virus {(SARS-CoV-2)}},
  author = {Lesley A. Shawler and Bryan Blair},
  journal = {Behav. Soc. Issues},
  volume = {30},
  number = {1},
  pages = {666--691},
  year = {2021}
}

@article{Markley2012,
  title = {Are gym surfaces reservoirs for {S}taphylococcus aureus? {A} point prevalence survey},
  author = {J. Daniel Markley and Michael B. Edmond and Yvette Major and Gonzalo Bearman and Michael Stevens},
  journal = {Am. J. Infect. Control.},
  volume = {40},
  number = {10},
  pages = {1008--1009},
  year = {2012}
}

@article{Kiborus2025,
  title = {Characterization of the Prevalence and Antibiotic Resistance of Staphylococcus Species in an Exercise Facility in {Central Kentucky}, {USA}},
  author = {Lilian Jeptoo Kiborus and S. Travis Altheide and Jason W. Marion},
  journal = {Hygiene},
  volume = {5},
  number = {1},
  pages = {2},
  year = {2025}
}

@article{Weitz2016,
  title = {An Oscillating Tragedy of the Commons in Replicator Dynamics with Game-Environment Feedback},
  author = {Weitz, Joshua S. and Eksin, Ceyhun and Paarporn, Keith and Brown, Sam P. and Ratcliff, William C.},
  year = 2016,
  month = nov,
  journal = {Proc. Natl. Acad. Sci. USA},
  volume = {113},
  number = {47},
  pages = {E7518--E7525}
}

@article{Tilman2020,
  title = {Evolutionary Games with Environmental Feedbacks},
  author = {Tilman, Andrew R. and Plotkin, Joshua B. and Akcay, Erol},
  year = 2020,
  month = feb,
  journal = {Nat. Commun.},
  volume = {11},
  pages = {915}
}

@article{Wang2020,
  title = {Eco-Evolutionary Dynamics with Environmental Feedback: {Cooperation} in a Changing World},
  author = {Wang, Xin and Fu, Feng},
  year = 2020,
  month = oct,
  journal = {EPL},
  volume = {132},
  pages = {10001}
}

@article{Chen2018,
  title = {Punishment and Inspection for Governing the Commons in a Feedback-Evolving Game},
  author = {Chen, Xiaojie and Szolnoki, Attila},
  year = 2018,
  month = jul,
  journal = {PLoS Comput. Biol.},
  volume = {14},
  pages = {e1006347}
}

@article{Smith1973,
  title = {The logic of animal conflict},
  author = {Smith, J Maynard and Price, G.},
  journal = {Nature},
  year = 1973,
  volume = {246},
  pages = {15--18}
}

@article{Friedman2010,
  title = {Gradient Dynamics in Population Games: {Some} Basic Results},
  author = {Friedman, Daniel and Ostrov, Daniel N.},
  year = 2010,
  month = sep,
  journal = {J. Math. Econ.},
  volume = {46},
  number = {5},
  pages = {691--707}
}

@article{Rosenthal1973,
  title = {A Class of Games Possessing Pure-Strategy {Nash} Equilibria},
  author = {Rosenthal, Robert W.},
  year = 1973,
  month = dec,
  journal = {Int. J. Game Theory},
  volume = {2},
  number = {1},
  pages = {65--67}
}

@article{Monderer1996,
  title = {Potential Games},
  author = {Monderer, Dov and Shapley, Lloyd S.},
  year = 1996,
  journal = {Games and Economic Behavior},
  volume = {14},
  number = {1},
  pages = {124--143}
}

@article{Taylor1997,
  title = {Evolutionary stability under the replicator and the gradient dynamics},
  author = {Peter Taylor and Troy Day},
  journal = {Evol. Ecol.},
  volume = {11},
  number = {5},
  pages = {579--590},
  year = {1997}
}

@book{Acemoglu2012,
  author    = {Daron Acemoglu and James A. Robinson},
  title     = {Why Nations Fail: {T}he Origins of Power, Prosperity, and Poverty},
  year    = {2012},
  month     = mar,
  publisher = {Crown Business},
  address   = {New York, NY}
}

@article{Bauch2003,
  title = {Group Interest versus Self-Interest in Smallpox Vaccination Policy},
  author = {Bauch, Chris T. and Galvani, Alison P. and Earn, David J. D.},
  year = 2003,
  month = sep,
  journal = {Proc. Natl. Acad. Sci. USA},
  volume = {100},
  number = {18},
  pages = {10564--10567}
}

@book{Strogatz:2024,
  title={Nonlinear Dynamics and Chaos},
  author={Strogatz, S. H.},
  year={2024},
  publisher={CRC Press},
  address={Boca Raton, FL}
}

@article{Schmid2021,
  author = {Laura Schmid and Krishnendu Chatterjee and Christian Hilbe and Martin A. Nowak},
  journal = {Nat. Hum. Behav.},
  title = {A unified framework of direct and indirect reciprocity},
  year = {2021},
  pages = {1292--1302},
  volume = {5}
}

@article{Hauser2009,
  author = {Hauser, Marc and McAuliffe, Katherine and Blake, Peter R.},
  journal = {Philos. Trans. R. Soc. Lond. B: Biol. Sci.},
  title = {Evolving the ingredients for reciprocity and spite},
  year = {2009},
  pages = {3255--3266},
  volume = {364}
}

@article{Trivers1971,
   author = {Trivers, Robert L.},
   title = {The Evolution of Reciprocal Altruism},
   journal = {Q. Rev. Biol.},
   volume = {46},
   pages = {35--57},
   year = {1971}
}

@article{Ohtsuki2006,
    title = {A simple rule for the evolution of cooperation on graphs and social networks},
    journal = {Nature},
    volume = {441},
    pages = {502--505},
    year = {2006},
    author = {Hisashi Ohtsuki and Christoph Hauert and Erez Lieberman and Martin A. Nowak}
}

@article{Lopez2025,
    title = {{SIR} models with vital dynamics, reinfection, and randomness to investigate the spread of infectious diseases},
    journal = {Commun. Nonlinear Sci. Numer. Simul.},
    volume = {140},
    pages = {108359},
    year = {2025},
    author = {Javier Lopez-de-la-Cruz and Alexandre N. Oliveira-Sousa}
}

@book{Nowak2006book,
    title={Evolutionary Dynamics: Exploring the Equations of Life},
    author={Nowak, M. A.},
    publisher={Harvard University Press},
    address = {Cambridge, MA},
    year={2006}
}

@incollection{brown1951fictitious,
  author    = {Brown, George W.},
  title     = {Iterative solution of games by fictitious play},
  booktitle = {Activity Analysis of Production and Allocation},
  editor    = {Koopmans, T. C.},
  pages     = {374--376},
  publisher = {Wiley},
  year      = {1951}
}

@article{gilboa1991social,
  author  = {Gilboa, Itzhak and Matsui, Akihiko},
  title   = {Social stability and equilibrium},
  journal = {Econometrica},
  volume  = {59},
  number  = {3},
  pages   = {859--867},
  year    = {1991}
}

@book{Smith1982,
    title = {Evolution and the Theory of Games},
    publisher = {Cambridge University Press},
    address = {Cambridge, UK},
    author = {Smith, John Maynard},
    year = {1982}
}

@book{fudenberg1998tlg,
  author    = {Fudenberg, Drew and Levine, David K.},
  title     = {The Theory of Learning in Games},
  publisher = {MIT Press},
  address = {Cambridge, MA},
  year      = {1998}
}

\clearpage

\renewcommand{\thefigure}{S\arabic{figure}}
\setcounter{figure}{0}

\renewcommand{\thetable}{S\arabic{table}}
\setcounter{table}{0}

\begin{figure}[t]
  \centering
  \begin{minipage}[b]{0.48\textwidth}
    \begin{overpic}[width=\textwidth]{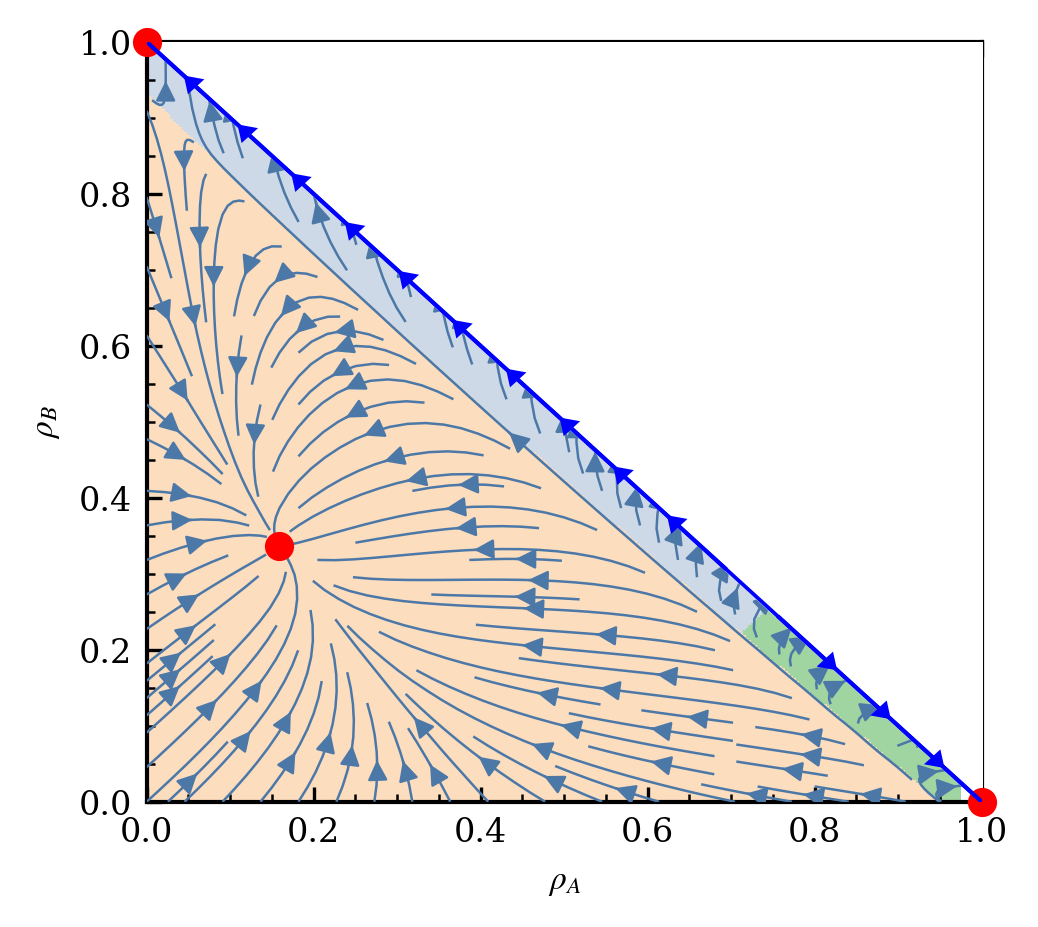}
            \put(4,94){\makebox(0,0)[l]{\footnotesize (a)}}   
    \end{overpic}
  \end{minipage}\hfill
  \begin{minipage}[b]{0.48\textwidth}
    \begin{overpic}[width=\textwidth]{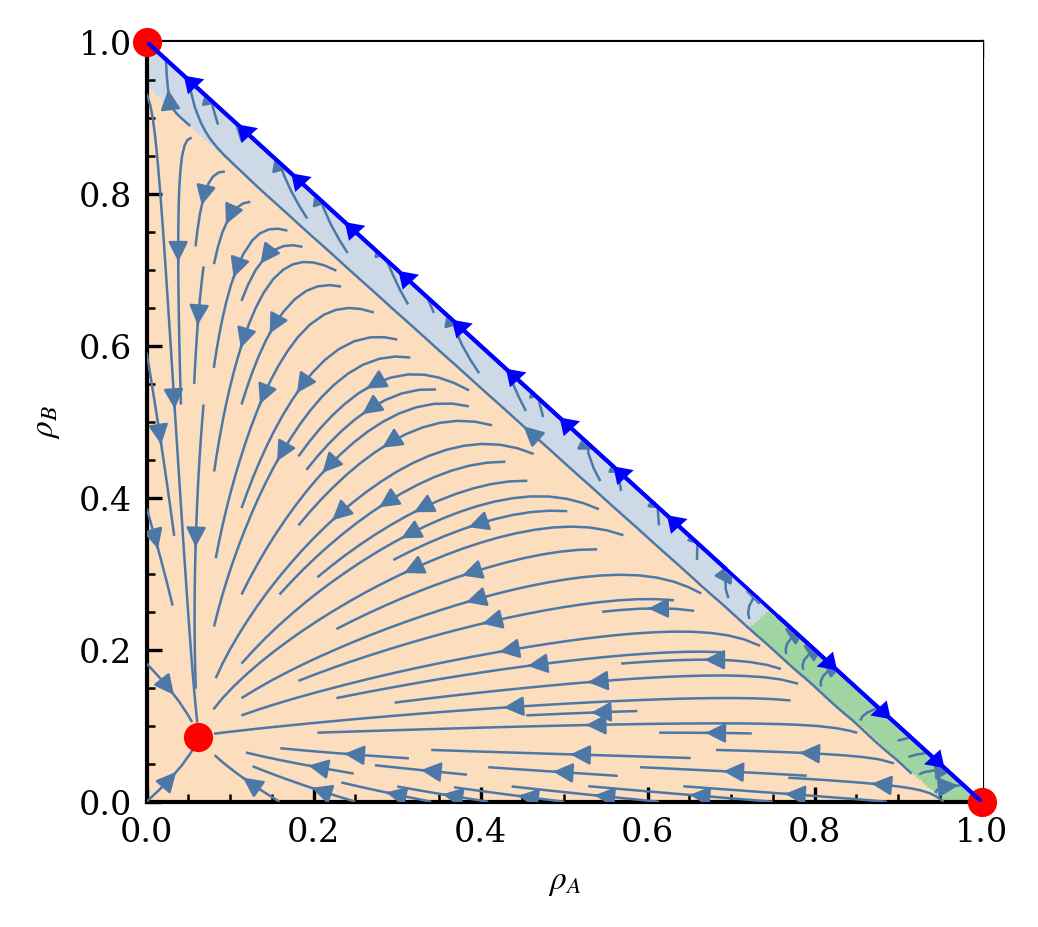}
            \put(4,94){\makebox(0,0)[l]{\footnotesize (b)}}   
    \end{overpic}
  \end{minipage}
  \caption{
    \textbf{Evolutionary dynamics and fixed point structure as a function of infection prevalence $x$.}
    (a) Same as Fig.~\ref{fig:combined_vector_fields}a, but with low infection prevalence, $x=0.2$.
    (b) Same as Fig.~\ref{fig:combined_vector_fields}a, but with high infection prevalence, $x=0.8$. 
}
  \label{fig:combined_vector_fieldsxdif}
\end{figure}

\end{document}